\documentclass[10pt,conference]{IEEEtran}
\usepackage[utf8]{inputenc}
\usepackage{todonotes}
\usepackage{tikz}
\usepackage[T1]{fontenc}
\usepackage[utf8]{inputenc}
\usepackage[font=small,labelfont=bf]{caption}
\usepackage{subcaption}
\usepackage{amsmath,amssymb,amsfonts}
\usepackage{graphicx}
\usepackage{color}
\usepackage{textcomp}
\usepackage{listings}
\usepackage{etoolbox}

\usepackage{graphicx}
\usepackage{multirow}
\usepackage{colortbl}
\usepackage{framed}
\usepackage{tabularx}
\usepackage[ruled]{algorithm2e} 
\usepackage{multirow,bigdelim}
\usepackage{enumitem}
\usepackage{balance}
\usepackage{url}
\usepackage[T1]{fontenc}
\usepackage[utf8]{inputenc}
\usepackage[font=small,labelfont=bf,tableposition=top]{caption}
\usepackage{booktabs}
\usepackage{threeparttable}

\makeatletter
\newcommand{\linebreakand}{%
  \end{@IEEEauthorhalign}
  \hfill\mbox{}\par
  \mbox{}\hfill\begin{@IEEEauthorhalign}
}
\makeatother

\begin{document}
\newcommand{\first}{project proposal}
\newcommand{\second}{progress report}
\newcommand{\numpull}{214}
\newcommand{\merged}{93}

\newcommand{\closed}{46}
\newcommand{\numproj}{24}

\newcommand{\numstudents}{154}
\newcommand{\numsurvey}{140}
\newcommand{\timec}{48}
\newcommand{\difficult}{47}

\newcommand{\sharing}{sharing}
\newcommand{\sharings}{sharings}
\newcommand*\circled[1]{\tikz[baseline=(char.base)]{
            \node[shape=circle,draw,inner sep=2pt] (char) {#1};}}
\newcommand{\faultloc}{query-based fault localization}
\newcommand{\faultloccap}{Query-based fault localization}
\newcommand{\cofind}{collaborative bug finding}
\newcommand{\rqcat}{RQ2}
\newcommand{\interrater}{0.76}
\newcommand{\studyrepair}{30\%}
\newcommand{\studycontext}{37\%}
\newcommand{\cofix}{collaborative bug fixing}
\newcommand{\driver}{open issue}
\newcommand{\precfix}{\textsc{Precfix}} 
\newcommand{\nav}{navigator}
\newcommand{\numstudy}{150} 
\newcommand{\numstudyproj}{150} 
\newcommand{\numevaljava}{152}
\newcommand{\numevaljavaproj}{25}

\newcommand{\numjavarel}{75}
\newcommand{\numandrel}{25}

\newcommand{\numevalandproj}{11}
\newcommand{\numevaland}{97}
\newcommand{\numiss}{10} 
\newcommand{\numeval}{249} 
\newcommand{\totalreport}{20} 
\newcommand{\numevalrepair}{50} 
\newcommand{\numreplypos}{six} 
\newcommand{\numevalreply}{seven} 
\newcommand{\numevalconfirm}{three} 
\newcommand{\numevalclose}{three}
\newcommand{\sizelowjava}{6} 
\newcommand{\sizehighjava}{31} 
\newcommand{\sizelowand}{6} 
\newcommand{\sizehighand}{31} 
\newcommand{\preckonelowjava}{0.3} 
\newcommand{\preckonehighjava}{1} 
\newcommand{\preckoneavgjava}{0.3399}
\newcommand{\preckoneavgand}{0.4433}
\newcommand{\preckoneavgjavagit}{0.3333}
\newcommand{\preckoneavgandgit}{0.4433}
\newcommand{\totalrelevant}{100}
\newcommand{\numcomment}{36}
\newcommand{\nummytoolrep}{10}
\newcommand{\nummytoolcon}{22}
\newcommand{\totalold}{10}
\newcommand{\pair}{navigator}
\newcommand{\numop}{five}
\newcommand{\numsim}{five}
\newcommand{\mytool}{GitHub-OSS Fixit} 
\newcommand{\Plus}{\mathord{\begin{tikzpicture}[baseline=0ex, line width=0.5, scale=0.06]\draw (1,0) -- (1,2);
\draw (0,1) -- (2,1);
\end{tikzpicture}}}
\newcommand{\Minus}{\mathord{\begin{tikzpicture}[baseline=0ex, line width=0.5, scale=0.06]
\draw (0,1) -- (2,1);
\end{tikzpicture}}}


\title{\mytool{}: Fixing bugs at scale in a Software Engineering Course}


\author{\IEEEauthorblockN{Shin Hwei Tan\IEEEauthorrefmark{1}, Chunfeng Hu\IEEEauthorrefmark{2}, Ziqiang Li\IEEEauthorrefmark{3}, Xiaowen Zhang\IEEEauthorrefmark{4} and Ying Zhou\IEEEauthorrefmark{5}}
\IEEEauthorblockA{
Department of Computer Science and Engineering, Southern University of Science and Technology, \\
Shenzhen, China\\
}
\IEEEauthorblockA{\IEEEauthorrefmark{1}tansh3@sustech.edu.cn; \IEEEauthorrefmark{2}hucf@sustech.edu.cn; \IEEEauthorrefmark{3}lizq2019@mail.sustech.edu.cn}
\IEEEauthorblockA{\IEEEauthorrefmark{4}zhangxw2019@mail.sustech.edu.cn;\IEEEauthorrefmark{5}zhouy8@mail.sustech.edu.cn}}

\maketitle

\begin{abstract}
Many studies have shown the benefits of introducing open-source projects into teaching Software Engineering (SE) courses. However, there are several limitations of existing studies that limit the wide adaptation of open-source projects in a classroom setting, including (1) the selected project is limited to one particular project, (2) most studies only investigated on its effect on teaching a specific SE concept, and (3) students may make mistakes in their contribution which leads to poor quality code. Meanwhile, software companies have successfully launched programs like Google Summer of Code (GSoC) and FindBugs ``fixit'' to contribute to open-source projects. Inspired by the success of these programs, we propose \mytool{}, a team-based course project where students are taught to contribute to open-source Java projects by fixing bugs reported in GitHub. We described our course outline to teach students SE concepts by encouraging the usages of several automated program analysis tools. We also included the carefully designed instructions that we gave to students for participating in \mytool{}. As all lectures and labs are conducted online, we think that our course design could help in guiding future online SE courses. Overall, our survey results show that students think that \mytool{} could help them to improve many skills and apply the knowledge taught in class. In total, \numstudents{} students have submitted \numpull{} pull requests to \numproj{} different Java projects, in which \merged{} of them have been merged, and \closed{} have been closed by developers. 
\end{abstract}




\section{Introduction}
\label{sec:intro}

Many researchers and educators have demonstrated success in adopting Open Source Software (OSS) into teaching Software Engineering (SE) courses~\cite{carrington2003teaching,openclass,bishop2016use,buchta2006teaching,petrenko2007teaching}. Compared to small programs written as part of a course assignment, OSS projects typically have large and realistic programs where their source code is publicly available under open-source licenses. Introducing OSS projects into SE courses are shown to provide several benefits, including offering chances to explore real world software~\cite{carrington2003teaching}, and to learn about coding practices~\cite{hu2018open}. 

However, there are several challenges and limitations of existing work that hinder the wide usage of OSS projects in a SE course. First, a study revealed several challenges in ensuring the diversity of the selected OSS projects when incorporating the selected projects into SE classes~\cite{hu2018open}. Indeed, most previous attempts in introducing OSS into SE curriculum usually focus on one particular OSS where the instructors have internal contacts with the development team of the selected OSS~\cite{hu2018open,ellis2007can}. Although having an on-site development team for the selected OSS means that students will get more guidance from the development team, this limits the diversity of the considered projects and restricts the freedom of students in selecting the projects in which they are interested in. Second, previous studies usually focus on a particular area (e.g., software design~\cite{carrington2003teaching} and evolution~\cite{buchta2006teaching}) but students can learn a lot of SE concepts and practices by participating in OSS projects. For example, according to prior mapping study~\cite{nascimento2013using}, there is rarely any study in areas such as configuration management, project management, and testing. Third, as students usually have little experience of contributing to OSS projects and have little exposure to the SE concepts and tools that may help them to excel in their projects, they tend to make some common mistakes in their contributions which may result in poor quality pull requests~\cite{hu2018open}. Some of these common mistakes include lack of tests, lack of comments and duplicated code.
\begin{table*}[!t]
\small
\centering
\caption{Comparison between existing approaches on OSS projects}
  \label{tab:compare}
\resizebox{\textwidth}{!}{%
\begin{tabular}{l|l|l|l|l}
\textbf{Approach}         & \textbf{Repository Selection}                                                                       & \textbf{Issue Selection}                                                                                   & \textbf{Participants}                                                                        & \textbf{Mentors}                                                                                                                                                                             \\\hline
\textbf{GSoC}             & \begin{tabular}[c]{@{}l@{}}Limited to organizations that \\ have signed up for GSoC\end{tabular} & \begin{tabular}[c]{@{}l@{}}Limited to the features selected\\by the mentoring organizations\end{tabular}           & \begin{tabular}[c]{@{}l@{}}Selected students that have \\ signed up for GSoC\end{tabular} & \begin{tabular}[c]{@{}l@{}}A few designated mentors from\\  each organization\end{tabular}                                                                                                   \\ \hline
\textbf{FindBugs ``fixit''} & Limited to FindBugs                                                                                 & Limited to issues in FindBugs                                                                               & \begin{tabular}[c]{@{}l@{}}Google software engineers\end{tabular}  & User who has reported the bug                                                                                                                                                               \\ \hline
\textbf{\mytool{}}   & Flexible. Can select any project                                                                    & \begin{tabular}[c]{@{}l@{}}Flexible. Can select issues that \\ fix bugs/implement a feature\end{tabular} & Students in a class                                                                          & \begin{tabular}[c]{@{}l@{}}User who has reported the GitHub\\ issue or the developer who \\ reviews the pull request.\end{tabular}
\end{tabular}}
\end{table*}


Meanwhile, software companies have started several projects in encouraging more students to participate in OSS projects. Since 2005, the Google Summer of Code (GSoC) program~\footnote{https://summerofcode.withgoogle.com/} has shown promising results in introducing students into OSS development by assigning the participants a programming project with an open-source organization during their summer break. The GSoC program offers many benefits, including (1) allowing students to get involved in open-source projects, (2) allowing students to practice Java programming skills, (3) providing opportunities to communicate with the mentors from open-source projects, etc. al. Meanwhile, in May 2009, Google has successfully conducted a large scale engineering effort (known as ``fixit'') where engineers reviewed thousands of warnings in FindBugs (a static analysis tool), repaired or filed bug reports against them~\cite{ayewah2010google}. 

Motivated by the success of GSoC and FindBugs ``fixit'', we introduce \mytool{}, a semester-long course project for a SE course where students select their favorite open-source Java projects and fix bugs reported in GitHub in these projects. Our key insight is that for each reported bug, there will be a person who filed the bug report, and he or she could serve as a virtual ``mentor''  who will give students the instructions required (e.g., the expected behavior) to reproduce the bug. Once a pull request was created for fixing the bug, the developer will play the role of a mentor by reviewing the pull request. Compared to GSoC, our \mytool{} project is more flexible because GSoC participants are required to register during a certain period of time and could only work with one OSS project that they have signed up for. In fact, the registration process takes long time (e.g., the GSoC 2020 program started accepting applications from the mentoring organizations since January 14 but students could only start coding on 1 June 2020), and students who have not been selected will not be able to participate in the program. To provide all students with the opportunities to contribute to OSS projects, we adapted the idea of fixing bugs from the FindBugs ``fixit'' project and brought it to life in a classroom setting. Table~\ref{tab:compare} shows the differences between our approaches, GSoC, and FindBugs ``fixit''. As shown in Table~\ref{tab:compare}, our \mytool{} is more general compared to the FindBugs ``fixit'' because students are free to choose their projects of interest instead of focusing only on FindBugs.

Overall, our contributions can be summarized as follows:
\begin{description}[leftmargin=*]
\item[Concept.]  We introduce \mytool{}, to the best of our knowledge, the first study that investigates on how to teach students to contribute to OSS projects by fixing bugs in their favorite Java projects. Our study exploits the fact that each GitHub issue has at least one mentor who will either provide students with more information to reproduce the bug or help them to review the pull request.
\item[Course outline and assignments.] To ensure the quality of written code and to increase the probability of the submitted pull requests being accepted, we describe how we design the course outline that focuses on using automated program analysis tools and applying XP practices during the bug fixing process. Due to COVID-19, our teaching is conducted online, which means that our current design can serve as guidance for future online courses. We introduced three assignments representing different stages of the \mytool{} project. In each stage, we provide detailed instructions which include steps for (1) project selection, (2) GitHub issues selection, (3) iteration planning, (4) assigning roles to the team members, (5) writing code according to the coding standards enforced via static analysis tools, (6) testing their implementations, and (7) making presentations to summarize their results. All teaching materials and data for the survey are publicly available at \url{https://github-fixit.github.io/}.
\item[Evaluation.] We evaluate the usefulness of \mytool{} by analyzing both qualitative and quantitative data. Specifically, in a survey participated by \numsurvey{} students, we evaluate whether \mytool{} helps them to improve their skills and SE knowledge taught in class. Although most students agreed that \mytool{} is time-consuming and challenging, most of them think that it helps them to improve many skills taught in class (e.g, XP practices, Git, GitHub, Java programming, etc.), and will recommend it for future course projects. We also analyze the statistics for the pull requests submitted by students. Overall, \numstudents{} students have submitted \numpull{} pull requests to \numproj{} different Java OSS projects where \merged{} have been merged, and \closed{} are awaiting developers' reply. 

\end{description}

\section{Background}
This section discusses the background that drives the design of \mytool{}, including (1) GitHub and GitHub classroom (the education technology used), and (2) eXtreme Programming (the software development methodology used).

\noindent \textbf{GitHub and GitHub classroom.}
\emph{GitHub}~\footnote{https://github.com/} is a software repository hosting service and a social coding platform. It leverages the git version control system and supports pull-based software development. \emph{GitHub classroom}~\cite{hsing2019using} is an open-source service launched by GitHub to facilitate using GitHub in the classroom setting. It supports several features, including importing class roster, online discussions, and automatic creation of software repositories for individual and group assignments. Internally, it uses an organization to store and manage course contents. We leverage the infrastructure supported by GitHub classroom to implement our \mytool{} project. GitHub uses Markdown~\footnote{https://guides.github.com/features/mastering-markdown/}, a lightweight syntax for formatting all texts in GitHub. Students in our class are required to submit answers for all assignments in the Markdown format. A GitHub \emph{issue} is a bug report written either by users or developers of an OSS project. A GitHub issue may be associated with a \emph{pull request} (a commit that fixes the reported bug or implements a specific feature). We designed \mytool{} based on the assumption that there are many open (unresolved) GitHub issues reported in OSS projects.
  
\noindent \textbf{eXtreme Programming (XP).}
EXtreme Programming (XP) is a software development methodology for small-to-medium sized teams developing software in the face of rapidly evolving requirements~\cite{beck2000extreme}. The key practices of XP include Test-driven development (TDD), iteration planning, coding standards, pair programming, and on-site customer~\cite{beck2000extreme}. \mytool{} aims to provide a platform for students to perform these practices in real world OSS projects. In XP, the customer is responsible for driving the project, giving project requirements and performing quality control via acceptance testing~\cite{martin2009role}. 

\begin{figure}[t!]
\centering
\includegraphics[width=\linewidth]{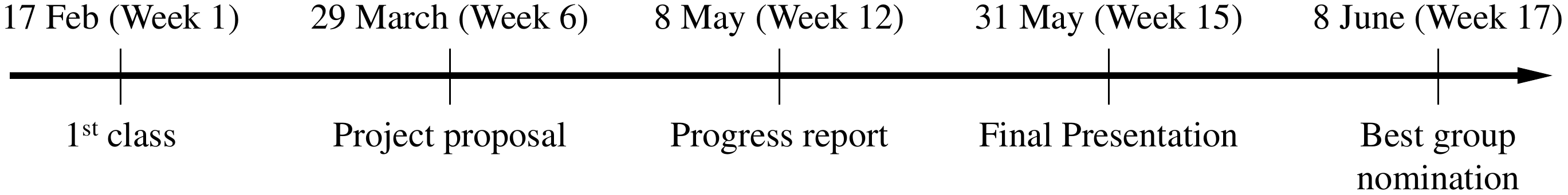}
  \caption{Timeline for the \mytool{} course project}
  \label{fig:timeline}
\end{figure}

\section{Course syllabus and setup}
We conducted \mytool{} as the course project for the Software Engineering course (CS304) in Southern University of Science and Technology (SUSTech). CS304 is a mandatory course for all computer science majors, and it is usually taken by students in the first semester of their junior year. In total, there are \numstudents{} junior students. The teaching staffs consist of two full-time instructors (one of them is the main instructor who conducts both lectures and labs, whereas another one is the lab instructor) and nine graders (seven of them are senior students and two of them are Master students). Although there are many teaching staffs, the instructors are responsible for giving lectures, whereas the graders are mainly in charge of (1) attending and grading the code review, and (2) marking lab attendances. 

\begin{table*}[]
\caption{Weekly schedule and topics covered during the lectures and the lab sessions}
  \label{tab:syllabus}
\centering
\resizebox{0.9\textwidth}{!}{%
\begin{tabular}{r|l|l}
Week & Lecture                                & Lab                                          \\\hline
1    & Why learn SE?                          & Roles in Software Development Teams          \\
2    & Version Control \& Build System        & Git and GitHub                               \\
3  & Waterfall model \& eXtreme Programming (XP) - Pair Programming & GitHub Classroom and Markdown                                                     \\
4    & Planning Game \& Unit Testing          & Unit testing with JUnit                      \\
5  & TDD \& Code Coverage                                           & Test coverage and Automated test generation with Evosuite                         \\
6    & Mutation Testing \& Evosuite (\emph{Project proposal due})         & PIT and Measuring Mutation Coverage          \\
7    & Reverse Engineering                    & Metrics in Java Projects                     \\
8    & -                                & -             \\
9  & Software Metrics                                               & Static Analysis Tools (PMD, Checkstyle, and Findbugs) \\
10   & Static Analysis                        & Reverse Engineering and Testing Android Apps \\
11   & Defensive Programming \& Documentation & Javadoc and Doxygen                          \\
12 & \emph{(Progress report due)} & -\\
13 & Software Reuse \& Component-based Software Engineering         & Code Review for Progress Report                                                   \\
14   & UI Design                              & Progress Report Presentation                 \\
15 & DevOps and Continuous Integration (\emph{Final presentation due})                             & Common vulnerabilities in Java programs                                            \\
16   & Security Engineering                   & Exam Review                                  \\
     &                                        &                                             
\end{tabular}}
\end{table*}

The entire course spans across 16 weeks with two hours of lectures and two hours of labs per week. Due to COVID-19, the entire course was conducted online with live teaching for both lectures and labs. To design relevant teaching materials that will help students in the \mytool{} project, the main instructor draws upon her prior experience of teaching SE related courses across three different universities and conducting research in SE. Table~\ref{tab:syllabus} shows the weekly schedule and the topics covered during the lectures and the labs. As shown in Table~\ref{tab:syllabus}, we taught various topics during the lectures and the labs (e.g., unit testing, software metrics, static analysis, reverse engineering, etc.), and introduced several relevant tools during the labs to help students in getting hands-on experience on the topics covered  (e.g., Git, Evosuite, PIT, MetricsReloaded Plugin, PMD, Checkstyle, and SpotBugs). By integrating various topics into the labs, we believe that this will help in avoiding the students' common mistakes when participating in OSS projects listed in previous study~\cite{hu2018open}. According to the study, students' common mistakes include lack of meaningful comments, lack of tests, bad naming, duplicated code, etc. To prevent students from making these mistakes, we explained the rules in writing good Javadoc comments in Week 11, emphasized the importance of testing in Week 4, and encouraged the use of static analysis tools for enforcing code standards in Week 9 and Week 10. To create a collaborative learning environment, instead of offering one-on-one help in tools installations, we encouraged students to post screenshots and error messages on the group chat or GitHub discussion to share common installation problems.

\section{The \mytool{} course project}
\label{sec:proj}
A prior study~\cite{muller2001case} revealed that small teams of less than eight people have less communication overhead and are more efficient than larger groups for adopting XP in a university environment. Hence, we asked all students to form a team of 5--6 people. Figure~\ref{fig:timeline} shows the timeline for the project. The \mytool{} project lasts for approximately 10 weeks (29 March--8 June 2020). The course project consists of three major stages: (1) project proposal, (2) progress report, and (3) final presentation. To encourage active participation, we announced at the start of the project that we will select one best group (students and teaching staffs votes for the best team) where each member will receive 5 bonus points. Students have around five weeks between the project proposal and the progress report stage, whereas they are given three weeks between the progress report and final presentation to complete the project. We allocate more time between the (1) and (2) stages compared to the duration between the (2) and (3) stages because the first step is always the hardest so students may need more time to get started with the OSS projects. 

We use the Research through Design (RtD)~\cite{frayling1993research,zimmerman2007research} approach to design and refine the instructions for each stage. Moreover, we gave students chances to practice writing documentation in the GitHub Markdown format by requiring that all answers to be written in the README.md file.

\subsection{Project Proposal}
In the project proposal, students need to perform several tasks, including (1) selection of OSS projects and their corresponding issues for fixing, (2) iteration planning for selecting the issues for each iteration, and (3) assigning roles for team members. For each task, we include the total score for answering each question accordingly. The total score for the project proposal is 20 points. The key idea behind the design of the project proposal is to give students the freedom to select the OSS projects where they will fix issues in the selected projects. As each GitHub issue $i$ is opened by a user or a developer of an OSS project (we call this person $p$), students could communicate with this person to get more information about $i$. As $p$ provides the project requirement (in the form of a GitHub issue), $p$ essentially plays the role of an XP customer. Moreover, the developer may perform code review for the GitHub issue $i$ if students make a pull request for $i$. 

\subsubsection{Selection of Projects and GitHub issues}
\label{sec:select}
Students need to select a list of open GitHub issues in 1--2 popular Java open-source projects. We select Java to be the target programming language because all students should have learned about some basic Java programming skills in a required class during the freshman year. Selecting a programming language where students have some familiarity will help to relieve the burden of learning a new language. To help students in project selection, we include a list of open-source Java projects~\footnote{https://github.com/dkorobtsov/automation-arsenal/blob/master/java/}. We selected this list because it categorizes projects based on their functionalities (e.g., commonly used utilities, IDE plugins, etc.) which makes it easier for students to select projects based on their interests. We also added several projects to the list (e.g., Soot~\cite{soot}, Randoop~\cite{randoop}, and JPF~\cite{jpf}) to encourage students to help in maintaining program analysis tools created by SE researchers.   

We list the criteria below for the selection of OSS projects: 
\begin{description}[leftmargin=*]
\item[Ease of use:] Each member of the team should be able to compile the Java project successfully on his or her computer without errors. If you select an app, the app should be able to be executed on a device (emulator/phone). 
\item[Existing Tests:] The project contains some test cases for checking for regression errors. 
\item[Popularity:] The number of stars in GitHub should be greater than 100.
\item[Actively Maintained:] There are recent commits (within a year) to the projects.
\item[Number of open GitHub issues:] The project should have at least 15 open GitHub issues that are bugs-related/feature-related (You need to check if the 15 GitHub issues are bugs/features instead of questions/documentations/tests). 
\item[Guidelines:] Check if the project has any contributing guidelines\footnote{usually available at https://github.com/a/b/blob/master/contributing.md where 'a/b' is the project name}. If yes, include its link. 
\end{description}

Specifically, we asked the students to check for the ``Ease of use'' and ``Existing Tests'' to ensure that they can compile and run the project successfully. To refrain students from selecting ``toy'' or outdated projects, we required students to check for the popularity of the project and whether the project is actively maintained. We required the students to select at least 15 open issues because ideally each student will fix two issues which sums up to 10--12 issues for a group of 5--6 students (additional issues serve as a backup for the scenario where some of these open issues have been closed by developers). We disallowed students to work on issues where users asked a question, or requested documentation/tests because they may not write code for fulfilling the requirements of such issues, and could not learn about the coding standards of OSS projects. Moreover, we refer students to the contributing guidelines of the GitHub project so that they read the guidelines before contributing. 

As selecting issues is the key step in the \mytool{} projects, we derive several criteria after referring to some open GitHub issues in OSS projects. Specifically, we include the following guidelines for selecting issues:
\begin{description}[leftmargin=*]
\item[Importance of the issue:] The selected issue should be important. For example, important issues have tags like ``up for grab'' or ``help wanted''. For most open-source projects, you can find the issues that are suitable for beginners under the ``contribute'' link (e.g, for INRIA/spoon, the ``contribute'' link is at https://github.com/INRIA/spoon/contribute).
\item[No fixing commit:] The issue should not have any fixing commits. Do not select an issue if a contributor has volunteered to fix it. 
\item[Reproducible:] For each selected issue, you need to make sure that it is reproducible and write a test case for reproducing it. Do not select an issue if (1) you do not understand the issue, or (2) someone mentioned that it is irreproducible. 
\item[Estimated lines of code to resolve issue:] The selected issue should be non-trivial to resolve. The estimated lines of code needed to resolve each issue should be greater than adding/modifying 10 lines of code. 
\item[Estimated time to resolve issue:] The estimated time taken to resolve one selected issue should be at least 1--2 weeks instead of 1--2 days.
\end{description}

The ``Importance of the issue'' criterion aims to ensure that students select non-trivial issues to fix, whereas the  ``No fixing commit'' criterion prohibits students from choosing issues that either have already been fixed or have pending pull requests. The ``Reproducible'' criterion is essential because students will need to write a test case to check for the validity and reproducibility of the selected issues. The ``Estimated lines of code to resolve issue'' and ``Estimated time to resolve issue'' are necessary to ensure that students will not select GitHub issues that do not take much time to fix. 



\subsubsection{Iteration Planning}
\label{sec:plan}
XP considers 1--3 weeks of development cycles as an iteration~\cite{beck2000extreme,beck2001planning}. As we have allocated six weeks for the entire project (until the final presentation), we planned two 3-week iterations to divide evenly the six weeks. In the project proposal, we listed the following instructions:
\begin{framed}
\noindent Select 5--20 issues to fix. Each issue needs to be selected by following the \textbf{guidelines for selecting issues}. The number of selected issues depends on (1) the number of group members, (2) the difficulty of the selected issues, and (3) the estimated time to fix the issue (there may be some related issues that could be fixed together so you will need to adjust the estimated time accordingly). Note that you need to consider the time taken in understanding the project and the issue (1 week for learning is allowed for each issue). The total estimated time for the group members should be at least 6 weeks per person. This means that if you have x team members in your group and the total estimated time for fixing all issues should be 6 multiplied by x (Total = 6x). For example, if an issue takes 3 weeks to fix, then each person could choose to fix 2 issues for the entire project. \textbf{Please plan your issues accordingly, fixing those issues that are more likely to be fixed by others first}. For all the selected issues, create a table such as below:
\resizebox{\linewidth}{!}{%
\begin{tabular}{|l|l|l|l|l|}
\hline
Link to issues &
  \begin{tabular}[c]{@{}l@{}}Type of \\ issues \\ (Bug/Feature)\end{tabular} &
  \begin{tabular}[c]{@{}l@{}}Estimated Time to \\ fix each issue\end{tabular} &
  \begin{tabular}[c]{@{}l@{}}Number of people\\ for fixing this issue\\ (\textless{}=2 members\\ for each issue)\end{tabular} &
  \begin{tabular}[c]{@{}l@{}}Estimated Difficulty \\(in a scale of 1-5,\\ 1 means very easy\\ to fix and 5 means\\ very difficult to fix)\end{tabular} \\\hline
https://... & Feature & 3 weeks & 1 & 5 \\\hline
...                                        &         &         &   &   \\\hline
\multicolumn{5}{|l|}{Total: 50 weeks}   \\\hline
\end{tabular}}
\end{framed}

\subsubsection{Roles for team members}
\label{sec:divide}
To ensure the diversity of team members and to encourage students to think about the roles in XP, we add the following requirement:
\begin{framed}
\noindent The team members should be diverse with each member playing different roles in a software development team. Include a division of roles for each team member for the first iteration of the project. Note that for each iteration, the member will get to take up different roles. \emph{(5 points)}
\resizebox{0.8\linewidth}{!}{%
\begin{tabular}{|l|l|}
\hline
Name & Role                                              \\\hline
     & Leader                                            \\\hline
     & Developer                                         \\\hline
     & Developer                                         \\\hline
     & Designer (Contribution guidelines \& code review) \\\hline
     & Tester                                            \\\hline
     & Developer \& Documentation (Javadoc)             \\\hline            
\end{tabular}}
\end{framed}

By assigning a leader to the project, the leader will ensure the progress of the team members. Moreover, we also assigned a tester for testing the code, and a designer who will check if the team members follow the contribution guidelines and perform appropriate code review. Meanwhile, we allocated a developer to be responsible for ensuring the quality of code comments. As stated in the instructions, we encourage students to play different roles for each iteration. 

\subsection{Additional Project Resources}
\label{sec:resource}
After the project proposal stage, students need to start working on fixing the GitHub issues. Around early April, the university has encouraged students to fill in a midterm feedback form to collect comments from students to ensure the quality of all classes that were offered online due to COVID-19. We received feedbacks from two students saying that the projects are too challenging, with one of these feedbacks requested us to provide more resources to discuss the steps involved in making the first contribution to an OSS project. Based on the students' feedback, we asked for help from two students with prior experience in contributing to OSS projects. One senior student (we will call this student $A$) had found a bug in an open-source Android app~\footnote{https://github.com/SimpleMobileTools/Simple-Calculator/issues/139} in the preceding semester (in a ``Software Testing'' class taught by the main instructor) and had fixed the bug where his pull request has been accepted by the developer. Based on his prior experience of fixing bug in an unfamiliar open-source app, student $A$ had recorded a video to share his experience. Another student who was a teaching assistant for the Software Engineering class (we will call this student $B$), has previous experience of contributing to the Kubernetes system~\footnote{https://github.com/kubernetes/kubernetes} so we have conducted an interview through an instant messaging app to request for his personal advice. After collecting information from both students, we posted several additional project resources: 
\begin{description}[leftmargin=*]
\item[R1:] A video on ``How to contribute to OSS project from 0''
\item[R2:] A picture that shows the interview Q\&A with student $B$
\item[R3:] An article on ``How to join the open-source community of Apache projects?'' (recommended by student $B$)
\item[R4:] Google Java Style Guide
\item[R5:] Android Resource Naming Cheat Sheet
\end{description}

Although sharing other external online resources may help to address the concerns raised by students, we asked for help from senior students because peer tutoring has shown to be effective in increasing learning for college students~\cite{falchikov2001learning}.

\subsection{Progress Report}
\label{sec:progress}
In the progress report, students need to (1) select the issues to implement for the current iteration, (2) write tests for the bug, (3) implement half of their selected issues (according to their iteration planning in Section~\ref{sec:plan}), (4) check their code against static analysis tools, (5) write Javadoc, (6) write JUnit tests for each added/modified public method, (7) plan for the remaining iteration, and (8) perform code review with a teaching assistant (TA). Specifically, we asked students to select the issues again (despite prior selection in the project proposal stage) to obtain the most up-to-date schedule for the current iteration. To assist students in planning, we asked students to select the issues for their first iteration based on either (1) importance of the issue, or (2) dependencies between issues; or (3) the ease of constructing tests. Moreover, we encourage test-driven development (TDD)~\cite{beck2003test} by asking students to write tests to reproduce the bugs before implemention. 

To ensure the quality of the submitted code, we asked students to run their code against three static analysis tools (i.e., Checkstyle~\cite{checkstyle}, SpotBugs~\cite{ayewah2008using}, and PMD~\cite{pmd}). We required students to change their code so that there is no serious error (severity=``error'') in Checkstyle. For SpotBugs, there should be no error with priority=``Medium'' and priority=``High'', whereas for PMD, there should be no error left when running against their submitted code. Moreover, we include a link to Google's coding standard for Javadoc~\cite{google-style} to encourage students to follow the guidelines for properly documenting their code. Moreover, students need to write at least two JUnit tests for each modified/newly added public method.
After the deadline of the progress report, the students performed a code review with a TA during the labs. During the code review, students need to follow the steps below: 

\begin{enumerate}[leftmargin=*]
\item The leader will first introduce their selected projects. Then, he will summarize the progress of the team in a few sentences and introduce the role of each team member.
\item The designer will explain the design of their code and describe the design plan for the next iteration. The designer will also show a demo of the implemented issues.

\item The TA will pick one person randomly to run the static analysis tool.

\item The person (developer) in charge of each issue will describe how it works. 

\item The person in charge of documentation will explain his implementation and the Javadoc comment for each method.

\item The testers will run all tests and shows the test results and code coverage results.

\item The team leader will end the code review by showing and explaining their plan for the next iteration. The TA will give some suggestions for future improvement.
\end{enumerate} 

We designed the above steps to ensure that all team members have chances to show the roles that they played during the code review. In step 3), we asked the TAs to select one person randomly among all team members to check whether all members are able to run and verify their code against the static analysis tools. In the preceding semester, we have conducted code review following similar steps. The key differences between the current code review and previous ones are: (1) the previous course project was on implementing an Android app from scratch instead of \mytool{}, and (2) the code review was conducted online through a video conferencing app instead of the face-to-face lab. During grading of the progress report, the TAs expressed their difficulties in distinguishing between the part of code that is written or modified by the students and those written by the developers. Moreover, we also notice that some students relied heavily on Evosuite, an automatic test generation tool introduced during Week 5 of the lab instead of writing tests manually.

\subsection{Final Presentation}
\label{sec:final}
\noindent \textbf{Deliverable.} For the final presentation, the main deliverable includes (1) source code, (2) tests, and (3) the slides for the presentation. Based on the feedbacks of the TAs, we added the requirement that students should mark the modified/added code with Javadoc comments using the format ``//CS304 Issue link: https://github.com/$a$/issues/123'' where $a$ is the name of the repository. Moreover, we also required students to mark the tests generated automatically by Evosuite to make it easier for the TAs to check if students have spent time in manual testing.

\noindent \textbf{Slides for presentation.} To encourage students to have good presentation skills and to ensure a fair grading scheme, we required all slides to have the content below:

\begin{itemize}[leftmargin=*]
\item Title slide with group name and students id
 
\item Introduction of the selected project(s) \textit{(5 points)}
\item Timeline of the project \textit{(2 points)}
\begin{itemize}
\item How well does your team follow the timeline? Explain with screenshots of version histories in GitHub                         \textit{(10 points)}

\end{itemize}
\item Screenshots and explanation of important issues \textit{(5 points)}
\item Testing techniques used and the number of tests \textit{(5 points)}
\item Screenshots to show results of static analysis tools \textit{(5 points)} 
\item Results showing the important issues implemented \textit{(2 points)}

\item Conclusions \textit{(1 point)}

\item Future Work (discuss lessons learned and future directions for your work) \textit{(5 points)}

\end{itemize}

We designed the above template based on our prior experience in making research presentations. An example presentation made by members of the best team is listed at \url{https://www.youtube.com/watch?v=EBcGYV51Np8}. 

\section{Evaluation}
We evaluate the effectiveness of \mytool{} using both qualitative and quantitative data. Specifically, our evaluation aims to address the following research questions:
\begin{description}
\item[Q1] What are the benefits and disadvantages of \mytool{} based on students' feedback?
\item[Q2] Do students perceive \mytool{} as a good course project?
\item[Q3] How many GitHub issues can the students resolve during the \mytool{} project?
\item[Q4] What is the effectiveness of the additional resources provided by senior students?
\end{description}


\noindent\textbf{Experimental Setup.}
To obtain the qualitative data, we offer extra credit to students who participated in a survey on the effectiveness of \mytool{}. Among the \numstudents{} taking the course, there are \numsurvey{} students who had participated in the survey. The survey contains a total of 12 questions where six of them asked the participants to rate items based on five-point Likert scales, three of them collects information about pull requests and code review, and two open-ended questions about the benefits and disadvantages of \mytool{}, and one question to collect basic information. 

\subsection*{Q1: Benefits and disadvantages of \mytool{}}

\begin{table}[t]
\small
\centering
\caption{Survey results for the question ``During the \mytool{}, I have improved my skills on:''}
  \label{tab:knowledge}
\begin{tabular}{l|r|r}
Skills & Mean & Standard Deviation\\ \hline
Version Control System & 4.26 & 0.75 \\
GitHub & 4.41 & 0.74 \\
Unit test & 4.31 & 0.74 \\
System/Integration tests & 3.65 & 0.98 \\
Code comment & 4.15 & 0.81 \\
Coding standard & 4.16 & 0.84 \\
Test-driven development & 3.83 & 0.93 \\
Java programming & 4.06 & 0.83 \\
Static Analysis tools & 4.32 & 0.78 \\
Teamwork & 4.03 & 0.92
\end{tabular}
\end{table}

\begin{table}[t]
\small
\centering
\caption{Survey results on the benefits of \mytool{}}
  \label{tab:benefits}
\resizebox{\linewidth}{!}{%
\begin{tabular}{l|r}
Benefits & \# of responses\\ \hline

Learn XP practices (coding standards, static analysis)                        & 71 \\
Teamwork                    & 57 \\
Read others' code & 55 \\
Practical project (can contribute to open-source)  & 54 \\
Learn GitHub and Git               & 50 \\
Improve skills (team communication, problem-solving)            & 45 \\ 
Realistic (learn about structure of big project) & 25 \\
Inherit the benefits of Git \& GitHub (efficient)   & 24 \\
Good experience (build confidence, fun, innovative) & 21 \\
Communicate with developer                    & 13 \\
Cannot find any                         & 5 \\
\end{tabular}
}
\end{table}

\begin{table}[!t]
\small
\centering
\caption{Survey results on the disadvantages of \mytool{}}
  \label{tab:cons}
\resizebox{\linewidth}{!}{%
\begin{tabular}{l|r}
Disadvantages & \# of responses \\ \hline

Cannot find any           & 52 \\
Time-consuming                   & \timec{} \\
Difficult/challenging            & \difficult \\ 
Hard to select suitable OSS projects/issues             & 42 \\ 
Hard to adjust/understand other's code      & 30 \\
Network connection problem (GitHub)             & 27 \\
Irresponsive developer             & 22 \\
Requires knowledge not taught in lecture/lab      & 20 \\
Imbalance level of difficulties in selected issues& 19 \\
Others & 19 \\
Hard to evaluate work done                 & 18 \\
Too rigid project requirement (e.g., only Java) & 18 \\
Problems with team cooperation      & 17 \\
Uncertainty and hard to plan & 14 \\
Issues with Git (branching and merging)                   & 4  \\
Problem with language (English)              & 3 \\
Some selected projects are not good & 3 \\
\end{tabular}
}
\end{table}

Our survey aims to thoroughly investigate the advantages and disadvantages of \mytool{}. To identify the skills that students may acquire when completing the \mytool{} projects, we evaluate whether the project helps in several aspects, including (1) version control system (Git), (2) GitHub (pull-based development), (3) unit test, (4) system/integration tests, (5) Code comment (Javadoc), (6) coding standards, (7) TDD, (8) Java programming, (9) static analysis tools, (10) teamwork. Table~\ref{tab:knowledge} shows the survey results for the question on the skills acquired by the students where each skill is rated on a five-point Likert scale, ranging from strongly disagree (score = 1) to strongly agree (score = 5). The first column in Table~\ref{tab:knowledge} shows the mean ratings selected by students, whereas the second column shows the standard deviation of the collected ratings for \numsurvey{} students (N = \numsurvey{}). Based on Table~\ref{tab:knowledge}, we can observe students generally agreed that they have observed improvement of several skills (mean = 4.41). Although it is expected that \mytool{} could help students in GitHub-related skills, we observe that most students ($\approx$89\%) either agree or strongly agree that \mytool{} helps them to learn about static analysis tools. We think that this improvement is due to our emphasis on the importance of using static analysis tools to check for the quality of the submitted code during the progress report and final presentation (Section~\ref{sec:progress}). Moreover, students also agreed that \mytool{} helps them to improve their Java programming skills (mean = 4.06).   

To gather more information about the benefits and disadvantages of \mytool{}, we asked students to list three benefits and three disadvantages in the survey. Then, we group together similar answers to analyze the results. Table~\ref{tab:benefits} shows the summarized results for the benefits of \mytool{}, whereas Table~\ref{tab:cons} shows the survey results for the disadvantages of \mytool{} (both tables are sorted in descending orders). As shown in Table~\ref{tab:benefits}, most students (71 responses) mentioned that they have learned about the key practices in XP (e.g., coding standards and static analysis). This is inline with the results in Table~\ref{tab:knowledge} where students agreed that they have improved skills on these XP practices. Moreover, students also appreciate the fact that they can improve their teamwork skills as they need to collaborate with their teammates for the project. Apart from the improvement in technical skills, we are also pleasantly surprised that 21 responses stated that \mytool{} provides a good overall experience (some students think that \mytool{} is fun and ``innovative''). 

In contrast, according to Table~\ref{tab:cons}, most students are positive about \mytool{} and could not find any disadvantages. Moreover, \timec{} students think that \mytool{} is time-consuming, whereas \difficult{} think that it is difficult and challenging.  Although we have provided detailed guidelines and additional resources (cf. Section~\ref{sec:resource}) for guiding students, 42 responses still expressed concerns in project and issue selection. Meanwhile, due to the COVID-19 pandemic, students need to participate in \mytool{} from home, and some of them (27 responses) mentioned that they have problems connecting to GitHub due to the slow connection. 

\subsection*{Q2: Suitability of \mytool{} as a course project}

We further evaluate whether students like the idea of having \mytool{} as a course project by asking them to rate in the five-point Likert scale for the question ``Overall, I would recommend \mytool{} for the class project'' with 1 being strongly disagreed and 5 being strongly agreed. Overall, most students either agree (42.14\%) or strongly agree (40\%) that they will recommend \mytool{} for the class project (mean = 4.16). Based on the survey results, we think that \emph{students like \mytool{} in general although they admitted that it is a time-consuming and challenging project}.

\subsection*{Q3: Statistics for the submitted pull requests} 
To gather quantitative data on the effectiveness of \mytool{}, we analyzed the submitted pull requests and the corresponding projects. 
\begin{table}[]
\small
\centering
\caption{Statistics on students' pull requests}
  \label{tab:pull}
\resizebox{\linewidth}{!}{%
\begin{tabular}{l|r}
Status & \# of pull requests\\ \hline
Merged      & \merged \\
Waiting for developer's reply        & \closed\\
Waiting for changes to pull requests   & 28 \\
Closed without explanation or fixed by developer                   & 24 \\
Developer abandoned the issue (won't fix) & 14 \\
Rejected pull requests                 & 9 \\\hline
Total & \numpull 
\end{tabular}}
\end{table}

\begin{table}[t]
\small
\centering
\caption{Statistics of the projects for the submitted pull requests}
  \label{tab:stat}
\resizebox{\linewidth}{!}{%

\begin{tabular}{l|r|r|r}
        & \# projects & \# of Stars in GitHub        & LOC           \\\hline
Java    & 19                   & 267--61422 & 4761--302644  \\\hline
Android & 5                    & 453--36739 & 10881--253578
\end{tabular}}
\end{table}

Table~\ref{tab:stat} shows the statistics of the projects in which students made at least one pull request. As shown in the table, the projects are quite diverse with 19 Java projects and five Android projects. To evaluate the quality of the submitted pull requests, we wrote a script that automatically checks the status (the status can be ``Open'', ``Closed'', and ``Merged'') and manually inspect those that are not merged. Table~\ref{tab:pull} shows the statistics of the submitted pull requests. In total, \numstudents{} students have submitted \numpull{} pull requests (1.39 pull requests on average). 
We think that the submitted pull requests have received positive feedback from developers because: (1) \merged{} of the submitted pull requests have been merged, and (2) 64.6\% of the survey participants responded that developers have performed code reviews for their pull requests (\emph{this confirms with our key observation that each GitHub issue has a designated mentor for ensuring code quality}). To further analyze the contribution by each team, we inspected the results for the best team (team JOJO). Five students from the team JOJO have successfully created 14 pull requests with 12 of them being merged. Overall, we consider \mytool{} a successful project as students were able to contribute positively to \numproj{} different OSS projects. 

\begin{table}[!t]
\small
\centering
\caption{The usefulness of the additional project resources}
  \label{tab:resource}
\begin{tabular}{l|r|r}
Resources     & Mean & Standard Deviation \\ \hline
R1: Video                        & 4.01 & 0.90               \\
R2: Q\&A Interview               & 3.90 & 0.88               \\ 
R3: Article & 3.94 & 0.91               \\ 
R4: Android naming convention              & 3.69 & 1.08               \\ 
R5: Google Java Style Guide           & 3.97 & 1.00               \\
\end{tabular}
\end{table}

\subsection*{Q4: Effectiveness of the additional project resources}
As described in Section~\ref{sec:resource}, we provided five resources (\textbf{R1}--\textbf{R5}) obtained from: (1) senior students who have prior experience in contributing to OSS projects, and (2) websites containing the naming conventions for Java and Android projects. In the survey, we asked students to rate the usefulness of these resources in five-point Likert scales (with 1 being strongly disagreed and 5 being strongly agreed). Table~\ref{tab:resource} shows the survey results on the effectiveness of the additional resources. Among all resources, students think that watching the video on ``How to contribute to OSS project from 0'' is the most useful (mean = 4.01). We think that the video is deemed useful because: (1) it is the most engaging among all resources, and (2) it is recorded by a student who had taken the class, which may create a sense of familiarity. Meanwhile, students rated the Google Java Style Guide (\textbf{R5}) as the second most effective project resources because we required students to follow the coding standards in \textbf{R5} if the selected projects have not specified their own coding standards. 

\section{Lessons Learned and Limitations}
\label{sec:lesson}
\noindent\textbf{OSS projects selection.} One of the key elements of \mytool{} is the chance for students to select their favorite OSS projects to work on. However, this poses several challenges. First, we need to carefully design the guidelines for the project selection to prevent students from selecting ``toy projects'' and to ensure that students could contribute to well-known OSS projects. Second, as shown in Table~\ref{tab:cons}, some students acknowledged the difficulties in selecting suitable OSS projects, whereas some students think that it would be hard to evaluate their work as the level of difficulties varies across projects and different issues. As it was our first time conducting \mytool{}, we rely on a list of OSS projects and a set of well-considered criteria for guiding the project selection. In subsequent semesters, we plan to share with students the list of \numproj{} OSS projects in which our students have submitted pull requests so that students could focus on fixing GitHub issues from these beginner-friendly projects (projects that have histories of accepting pull requests created by our students).   

\noindent\textbf{GitHub issues selection.} Many students expressed difficulties in selecting GitHub issues as the complexity and time required for fixing a bug may vary substantially across different projects. In general, it is not possible to control the complexity of the selected issues, so we could provide guidance by recommending students to focus on issues with labels such as ``good first issue'' and ``help wanted''. Moreover, some students mentioned that the GitHub issues selected during the project proposal stage were fixed before they started working on these issues. To solve this problem, during the lecture and in the video ($R1$), we advocated following a TDD approach where students create tests to reproduce the bug after selecting the issues. To verify the validity of the selected issues, we also recommended students to communicate with the developers by leaving comments to express the intention to contribute. 

\noindent\textbf{\mytool{} as course project.} Table~\ref{tab:cons} shows that most students think that \mytool{} is time-consuming and challenging. This feedback is expected because students taking the class are all beginners who do not have much experience contributing to OSS projects. Based on the midterm feedback, we include more resources to help students to overcome the barriers of getting started (c.f. Section~\ref{sec:resource}). As most students agreed that the extra resources are useful, we plan to release these resources earlier in subsequent semesters. 

\noindent\textbf{Project requirements.} In the survey, there are 18 responses that complained that the project requirement is too rigid. Specifically, four responses mentioned that they could only select Java projects, whereas other students mentioned that they need to write unnecessary Javadoc comments or spend time in writing the answers for the progress report and the final presentation. We require all selected OSS projects to be written in Java because (1) all tools taught during the labs could only analyze Java programs, and we would like to encourage the usages of these tools, and (2) it is harder to assess the relative progress of each team if the code is written in different programming languages. Despite this restriction, we think that the \numproj{} selected projects are still quite diverse. Examples of the selected projects include the Guice framework by Google~\footnote{https://github.com/google/guice}, Fastjson library by Alibaba~\footnote{https://github.com/alibaba/fastjson}, and the Nextcloud Notes app~\footnote{https://github.com/stefan-niedermann/nextcloud-notes}. To allow students to select projects written in different languages, instructors need to collect more resources about the installations of automated tools. Another solution would be to rely on students to search for the relevant tools but this process may incur additional burdens to students. 

\noindent\textbf{Grading.} Although there are 18 responses saying that it is hard to evaluate the work done (Table~\ref{tab:cons}), we had not received any personal complaints about the grading being unfair or inconsistent. We attribute this success to our detailed instructions and our transparent grading scheme (as shown in Section~\ref{sec:proj}, we have clearly listed the assigned score for each question). In fact, we received a personal positive feedback from a student who said that she appreciated the fact that we did not deduct any points for pending pull requests (this helps to mitigate the problem of irresponsive developers).     

\noindent\textbf{Quality of pull requests.} As the teaching staffs do not have experience in developing most of the selected projects, we cannot perform extensive code reviews for the submitted pull requests. To ensure the quality of the submitted pull requests, we advocate the enforcement of coding standards, the usage of static analysis tools, testing and writing documentation. We assume that by following strictly the basic Software Engineering concepts and the key practices of XP, the submitted pull requests will have higher chances of getting accepted by developers. This assumption is inline with previous study that taught Software Engineering with limited resources~\cite{6227025}. The fact that \merged{} of the pull requests submitted by our students are merged by developers confirmed with this observation. In the future, it is worthwhile to study the effects of introducing code review bots for reviewing pull requests made by students~\cite{wessel2020effects}. 

\section{Related Work}

\noindent\textbf{Teaching XP practices.} There are several previous studies on adopting XP practices for teaching programming~\cite{keefe2006adopting,pancur2003towards}. Although we designed our teaching materials based on XP and its key practices, our goal is to ensure that the quality of the code written is acceptable by developers of OSS projects.  

\noindent\textbf{Studies on GitHub.} Several studies have shown the effectiveness of using GitHub Classroom for teaching~\cite{fiksel2019using,griffin2013github,feliciano2016student,tan2020collaborative}. Different from all these studies, \mytool{} focuses on teaching students how to contribute to OSS projects by creating good quality pull requests. 

\noindent\textbf{Studies on adopting OSS projects.} Many prior studies suggested adopting OSS projects into teaching software engineering courses~\cite{carrington2003teaching,openclass,bishop2016use,buchta2006teaching,petrenko2007teaching,hu2018open,pinto2019training,tafliovich2019teaching}. These studies usually assume that the instructors have pre-selected several projects for all students. Meanwhile, we include the instructions for selecting suitable OSS projects to provide students with the freedom to choose their projects of interest. We have not found any other study that offers this freedom. 

To encourage contributions to OSS, DigitalOcean organized Hacktoberfest~\footnote{https://hacktoberfest.digitalocean.com/} where those who submit pull requests on October received free t-shirts but it is not an education-focus program. Similar to GSoC~\cite{silva2017long,silva2020google}, our \mytool{} project focuses on encouraging participate in OSS projects. Our project differs from GSoC because: (1) our project does not require pre-registration which makes it more accessible and easier to incorporate into a classroom setting than GSoC, and (2) GSoC have appointed mentors for giving advice to students, whereas the person who has opened an issue will provide advice due to the need to resolve the issue. 

\noindent\textbf{Related studies on bug fixing.} Many prior approaches and studies proposed applying automated program repair techniques in fixing introductory programming assignments~\cite{gulwani2018automated,yi2017feasibility,parihar2017automatic}. Orthogonal to these studies, our approach focuses on teaching students how to contribute to OSS projects by making pull requests manually to fix bugs. In the future, it is worthwhile to study the effect of apply automated repair techniques in fixing bugs for \mytool{}.

\section{Conclusion}
We introduce \mytool{}, a team-based project that encourages contributions to OSS development by having students fix bugs in their selected projects. We design \mytool{} based on the insight that each GitHub issue has a designated reporter who can give more information about the bug or feature and a developer who can help in reviewing the pull requests. This means that each GitHub issue has at least one mentors who will guide students. This key insight helps us to eliminate the needs for the designated mentors and the long registration process in GSoC. We also described how we designed the course outline and the detailed instructions. We evaluate the effectiveness of our approach qualitatively and quantitatively. In the survey, students have provided positive feedbacks on the usefulness of \mytool{} in helping them to improve their skills and apply the knowledge taught in class. Overall, students have submitted \numpull{} pull requests where \merged{} are merged. 
Although this was the first time where we have conducted \mytool{}, we think that it is quite successful considering students' feedback and the relatively large number of submitted pull requests. As we conducted all lectures and labs online, our course outline, teaching materials, and instructions for \mytool{} can help to guide the design for future online SE courses. In the future, we plan to continue using \mytool{} as a semester-long project for SE course, and refine it based on our findings in Section~\ref{sec:lesson}.    






\section*{Acknowledgments}
We thank all SUSTech students taking CS304 (Spring 2020) for their participations. This work was partially supported by the Natural Science Foundation of Guangdong Province (Grant No. 2020A1515011494). We thank the Open Source Competence Center in Huawei for sponsoring our future projects. As this project won the World Teacher Day Challenge, we thank the organizers for the award.

\balance
\bibliographystyle{IEEEtran}
\bibliography{bibliography}



\end{document}